\documentclass[aps,pre,showpacs,twocolumn,floats]{revtex4}
\usepackage{amssymb}

\usepackage{amsmath}
\usepackage{graphicx,psfig}
\usepackage{dcolumn}
\usepackage{bm}

\input{tcilatex}

\begin{document}

\title{Two-phonon spin-lattice relaxation of rigid atomic clusters}
\author{C. Calero, E. M. Chudnovsky, and D. A. Garanin}
\affiliation{Department of Physics and Astronomy, Lehman College,
City University of New York \\ \mbox{250 Bedford Park Boulevard
West, Bronx, New York 10468-1589, U.S.A.}}
\date{\today}

\begin{abstract}
Spin-phonon relaxation due to two-phonon processes of a spin
cluster embedded in an elastic medium has been studied. For the
case of uniaxial anisotropy, relaxation rates due to Raman
processes and processes involving the emission of two phonons have
been obtained. For a biaxial spin Hamiltonian, the rates of
transitions between tunnel-split levels have been computed. By
comparison with the rates of corresponding direct processes, we
have established temperature ranges where the Raman mechanism
dominates over the one-phonon decay mechanism.

\end{abstract}

\maketitle
\section{Introduction}
Relaxation of spins in a paramagnetic solid is a problem of
fundamental interest. It is also related to important applications
such as spin resonance and the use of spin as a qubit. In a
paramagnet, the decay of a spin state is due to the interactions
with phonons, nuclear spins, dipolar fields, etc. In principle,
one could suppress the interaction with nuclear spins and dipolar
fields but the interaction with phonons always remains. Often it
dominates spin relaxation. Thus, spin-lattice interactions provide
the most fundamental upper bound for the lifetime of spin states
in paramagnets. The spin relaxation due to interaction with the
lattice can occur by various mechanisms. The most studied and
often the dominant ones are direct processes, in which a single
quantum is exchanged between the spin system and the lattice. It
was pointed out by Waller in studying the modulation of the
spin-spin interaction by the lattice waves that, unlike in
electromagnetic phenomena, the inelastic scattering of a phonon
combined with a transition in the spin system could be very
important \cite{Waller}. This is a two-phonon process consisting
of the absorption of one phonon and the emission of another phonon
with different frequency. The mechanism is analogous to the Raman
effect in optical spectroscopy and it is often referred to as a
Raman process. Spin-lattice relaxation mechanisms based on dipolar
interactions (Waller) were insufficient, however, to account for
the transition rates measured in experiment. Heitler and Teller
\cite{Heitler} considered Raman processes based on a more potent
mechanism based upon modulation of the crystal electric field
under the action of the lattice vibrations. Their theory was
further developed by Kronig \cite{Kronig} and Van Vleck
\cite{VanVleck}. They obtained a spin-phonon coupling based on the
spin-orbit interaction that permitted calculation of relaxation
rates that were of the same order of magnitude as the experimental
ones. Later on, Orbach \cite{Orbach} managed to simplify the
treatment of the problem by expanding the crystal electric
potential in powers of the fluctuating strain caused by the
lattice vibrations.

Recently, this problem has received new attention in connection
with spin relaxation of molecular clusters. The spin of many such
clusters is formed inside a relatively rigid magnetic core that
can rotate in the presence of the deformation field but is more
resistant to distortions of the core itself. It has been noticed
that the spin relaxation of such a cluster can be obtained within
a model that is parameter free, that is, it gives the relaxation
rates in terms of the known crystal field Hamiltonian of the
magnetic core $\hat{\cal{H}}_A$ \cite{CGS}. Even for non-rigid
clusters, calculation of the effect of rotations is meaningful. It
has been theoretically established that spin-phonon relaxation
rates due to both, one-phonon and multi-phonon processes, are
inversely proportional to some high powers of the sound velocity
\cite{A&B}. Since longitudinal phonons have a larger sound
velocity than the transverse phonons, processes involving
longitudinal phonons can be safely neglected. The effects of the
transverse phonons can be split into shear deformations of the
lattice cell and local rotations of the lattice that preserve the
symmetry of the crystal field. To describe deformations of the
first kind one needs to employ terms in the Hamiltonian containing
phenomenological coupling constants, whereas the local rotations
can be described by a parameter-free spin-phonon Hamiltonian that
is defined solely by the form of $\hat{\cal{H}}_A$. In general,
processes due to the shear distortion of the lattice and those due
to the local rotation of the lattice should result in comparable
relaxation rates. Even in this case, the latter are of a
fundamental importance because they provide a parameter-free lower
bound on the decoherence of any spin-based qubit. For rigid spin
clusters, interaction of the spin with rotations of the crystal
field is the only source of spin-lattice relaxation.

The angle of rotation of the crystal field axes $({\bf e}^{(1)},
{\bf e}^{(2)}, {\bf e}^{(3)})$ due to the action of the phonon
field ${\bf u}({\bf r})$ is given by
\begin{equation}\label{angle}
\delta {\bf \phi}({\bf r}) = \frac{1}{2} \nabla \times {\bf
u}({\bf r}).
\end{equation}
Because $\mathcal{\hat{H}}_A$ is a scalar, the rotation of $({\bf
e}^{(1)}, {\bf e}^{(2)}, {\bf e}^{(3)})$ is equivalent to the
rotation of the vector ${\bf S}$ in the opposite direction. As it
is known, the rotation of the operator ${\bf S}$ can be performed
by the $(2S+1)\times(2S+1)$ matrix in the spin space,
\begin{equation}\label{spin-rotation}
{\bf S} \rightarrow \hat{R}{\bf S}\hat{R}^{-1}, \qquad \hat{R} =
e^{-i{\bf S}\cdot \delta{\bf \phi}}\,.
\end{equation}
Then, the total Hamiltonian in the presence of phonons can be
written in the form \cite{CGS}
\begin{equation}\label{total Hamiltonian}
{\mathcal{\hat{H}}} = \hat{R}{\mathcal{\hat{H}}}_A\hat{R}^{-1} +
{\mathcal{\hat{H}}}_Z + {\mathcal{\hat{H}}}_{ph}\,,
\end{equation}
where ${\mathcal{\hat{H}}}_A$ is the crystal-field Hamiltonian in
the absence of phonons, ${\mathcal{\hat{H}}}_Z$ is the Zeeman
Hamitonian and ${\mathcal{\hat{H}}}_{ph}$ is the Hamiltonian of
harmonic phonons. In these formulas $\delta {\bm \phi}$ must be
treated as an operator. The canonical quantization of phonons and
Eq.\ (\ref{angle}) yield
\begin{equation}\label{angle2}
\delta \bm {\phi }  =  \sqrt{\frac{\hbar }{8\rho V}}\sum_{\mathbf{k}%
\lambda }\frac{\left[ i\mathbf{k}\times \mathbf{e}_{\mathbf{k}\lambda }%
\right] e^{i\mathbf{k\cdot r}}}{\sqrt{\omega _{\mathbf{k}\lambda
}}}\left( a_{\mathbf{k}\lambda }+a_{-\mathbf{k}\lambda }^{\dagger
}\right)
\end{equation}
where $\rho$ is the mass density, $V$ is the volume of the
crystal,
 $\mathbf{e}_{\mathbf{k}\lambda }$ are unit polarization vectors, $
\lambda =t_{1},t_{2},l$ denotes polarization, and $\omega
_{k\lambda
}=v_{\lambda }k$ is the phonon frequency.\\

For one-phonon processes, relaxation rates of spin-phonon
transitions have been recently computed with the help of the above
formalism in Ref. \onlinecite{CGS}. Such processes dominate
spin-phonon relaxation at zero temperature when no thermal phonons
are present in the system. At finite temperature, however,
two-phonon processes may take over. In this paper we study two
kinds of two-phonon processes. The first kind consists of an
inelastic scattering of phonons by the spin-system, or spin-phonon
Raman processes. It corresponds to the annihilation of an incoming
phonon of frequency $\omega_{\bf k}$ and the creation of an
outgoing phonon of frequency $\omega_{\bf q}$, with $ {\bf k}$ and
${\bf q}$ being the corresponding wave vectors. The second kind
involves emission of two phonons. Note that the conservation of
the energy in spin-phonon interactions requires
\begin{equation}\label{energycons}
\hbar \omega_{\bf q} \pm \hbar \omega_{\bf k} = \Delta E >0\,,
\end{equation}
where $\Delta E$ is the energy difference between the spin-states
and $\omega_{\bf k}, \omega_{\bf q}$ are the frequencies of the
phonons involved in the transition. The plus sign applies to
processes involving the emission of two phonons and the minus sign
applies to Raman processes. When $\Delta E \ll \hbar \omega_D$
(with $\omega_D$ being the Debye frequency), Eq.\
(\ref{energycons}) causes the phase space of phonons to be much
greater for Raman processes than for processes involving the
emission of two phonons. Consequently, the spin-phonon Raman
scattering usually dominates over the processes involving the
emission of two phonons. However, the same condition
(\ref{energycons}) implies that the energy of the phonon emitted
in the Raman scattering process, $\hbar \omega_{\bf q}$, must
satisfy $\hbar \omega_{\bf q}> \Delta E$, whereas in the process
involving the creation of two phonons their energy must be smaller
than $\Delta E$. Therefore, if $k_BT \ll \Delta E$, the total
number of phonons available to carry out the spin transition may
be much greater in the two-phonon emission case than in the Raman
case, so that the former case can become dominant. In both cases
the matrix element of the transition is a sum of two terms. The
first term, $M^{(2)}$, comes from the first order of the
perturbation theory on the spin-phonon coupling containing a
product of two phonon displacement fields. The second term,
$M^{(1+1)}$, comes from the second order of the perturbation
theory on the spin-phonon coupling that is linear in the phonon
displacement field. In some cases these two terms interfere, so
that the resulting transition rate, based upon $|M^{(2)} +
M^{(1+1)}|^2$, is different from the one obtained by adding up the
rates, $|M^{(2)}|^2$ and $|M^{(1+1)}|^2$, that each term would
produce by itself, as was incorrectly done in the past \cite{A&B}.

\section{Matrix elements of two-phonon processes}

The treatment of two-phonon processes requires consideration of
terms up to second order in phonon amplitudes in the Hamiltonian:
\begin{equation}\label{angle-expansion}
\hat{R}{\mathcal{\hat{H}}}_A\hat{R}^{-1} \simeq
{\mathcal{\hat{H}}}_A + {\mathcal{\hat{H}}}_{\mathrm{s-ph}} \qquad
{\mathcal{\hat{H}}}_{\mathrm{s-ph}}  =
{\mathcal{\hat{H}}}_{\mathrm{s-ph}}^{(1)} +
{\mathcal{\hat{H}}}_{\mathrm{s-ph}} ^{(2)}\,,
\end{equation}
with
\begin{eqnarray}\label{Hsph}
{\mathcal{\hat{H}}}_{\mathrm{s-ph}}^{(1)} & = &
i\left[{\mathcal{\hat{H}}}_A, {
S_{\alpha}}\right]\,\delta\phi_{\alpha} \nonumber \\
{\mathcal{\hat{H}}}_{\mathrm{s-ph}}^{(2)} & = &
\frac{i^2}{2!}\left[\left[{\mathcal{\hat{H}}}_A,
S_{\alpha}\right],S_{\beta}\right]\delta\phi_{\alpha}\delta\phi_{\beta}\,.
\end{eqnarray}
The total Hamiltonian can then be written as
\begin{equation}\label{total Hamiltonian2}
{\mathcal{\hat{H}}} = {\mathcal{\hat{H}}}_0  +
{\mathcal{\hat{H}}}_{\mathrm{s-ph}}\,,
\end{equation}
where ${\mathcal{\hat{H}}}_{0}$ is the Hamiltonian of
non-interacting spin and phonons,
\begin{equation}\label{non-int-Hamiltonian}
 {\mathcal{\hat{H}}}_0  = {\mathcal{\hat{H}}}_S + {\mathcal{\hat{H}}}_{\mathrm{ph}}\,,
\end{equation}
and
\begin{equation}\label{spin-Hamiltonian}
{\mathcal{\hat{H}}}_S = {\mathcal{\hat{H}}}_A +
{\mathcal{\hat{H}}}_Z\,,
\end{equation}
is the spin Hamiltonian. \\
We will study spin-phonon transitions between the eigenstates of
${\mathcal{\hat{H}}}_0$, which are direct products of the spin and
phonon states,
\begin{equation}\label{eigenstates}
\left| \Psi _{\pm }\right\rangle =\left| \psi _{\pm }\right\rangle
\otimes \left| \phi _{\pm }\right\rangle .
\end{equation}
Here $\left| \psi _{\pm }\right\rangle $ are the eigenstates of
$\mathcal{\hat{H}}_{S}$ with eigenvalues $E_{\pm }$
($E_{+}>E_{-}$) and $\left| \phi _{\pm }\right\rangle $ are the
eigenstates of $\mathcal{\hat{H}}_{\mathrm{ph}}$ with energies
$E_{\mathrm{ph},\pm }$. Spin-phonon transitions conserve energy
\begin{equation}
E_{+}+E_{\mathrm{ph},+}=E_{-}+E_{\mathrm{ph},-}.
\label{Energyconservation}
\end{equation}
To obtain the relaxation rate of the transition $\left| \Psi _{+
}\right\rangle \rightarrow \left| \Psi _{- }\right\rangle$ one
needs to evaluate the matrix element of the process. This matrix
element is the sum of the matrix element with $
{\mathcal{\hat{H}}}_{s-ph}^{(2)} $ and that with $
{\mathcal{\hat{H}}}_{s-ph}^{(1)}$ in the second order
\cite{Sakurai}:
\begin{equation}\label{M}
M = M ^{(2)} + M^{(1+1)}\,,
\end{equation}
where
\begin{equation} \label{M2}
M^{(2)}=\left\langle \Psi _{-}\left|
{\mathcal{\hat{H}}}_{\mathrm{s-ph}}^{(2)}\right| \Psi
_{+}\right\rangle
\end{equation}
and
\begin{eqnarray}
M^{(1+1)} &=&\sum_{\xi }\frac{\left\langle \Psi _{-}\left|
{\mathcal{\hat{H}}}_{\mathrm{s-ph}}^{(1)}\right| \Psi _{\xi
}\right\rangle \left\langle \Psi
_{\xi }\left| {\mathcal{\hat{H}}}_{\mathrm{s-ph}}^{(1)}\right| \Psi _{+}\right\rangle }{%
E_{+}-E_{\xi }}. \nonumber \\ \label{M11}
\end{eqnarray}
Here $\xi $ labels intermediate spin-phonon states.

\section{Raman processes}

For the Raman processes of interest, a phonon with the wave vector
$\mathbf{k}$ is absorbed and a phonon with the wave vector
$\mathbf{q}$\ is emitted. We will use the following designations
for the phonon states
\begin{equation}
\left| \phi _{+}\right\rangle \equiv \left| n_{\mathbf{k}},n_{\mathbf{q}%
}\right\rangle ,\qquad \left| \phi _{-}\right\rangle \equiv \left| n_{%
\mathbf{k}}-1,n_{\mathbf{q}}+1\right\rangle  \label{phiviankRaman}
\end{equation}
In this case, the conservation of the energy reads:
\begin{equation}
\Delta E = E_{+}- E_-=\hbar \omega_{\bf q} - \hbar \omega_{\bf k}.
\label{EnergyconservationRamanE2}
\end{equation}

The matrix element of the Raman process can be written as $ M_R =
M_R^{(2)} + M_R^{(1+1)}$. According to equations (\ref{Hsph}) and
(\ref{M2})
\begin{eqnarray}
M_{R}^{(2)} =&-&\frac{1}{2}\left\langle \psi _{-}\left| \left[
\left[ {\mathcal{\hat{H}}}_{A},S_{\alpha }\right] ,S_{\beta
}\right] \right| \psi _{+}\right\rangle
\notag \\
& \times & \left\langle n_{\mathbf{k}}-1,n_{\mathbf{q}}+1\left|
\delta
\phi _{\alpha }\delta \phi _{\beta }\right| n_{\mathbf{k}},n_{\mathbf{q}%
}\right\rangle .  \label{MR2b}
\end{eqnarray}
It is convenient to express the phonon matrix element as
\begin{equation}
\left\langle n_{\mathbf{k}}-1,n_{\mathbf{q}}+1\left| \delta \phi
_{\alpha
}\delta \phi _{\beta }\right| n_{\mathbf{k}},n_{\mathbf{q}}\right\rangle =M_{%
\mathrm{ph-R}}^{\alpha \beta }+M_{\mathrm{ph-R}}^{\beta \alpha },
\label{MphRDef}
\end{equation}
where
\begin{equation}\label{MphRDef2}
M_{\mathrm{ph-R}}^{\alpha \beta } =\left\langle
n_{\mathbf{q}}+1\left|
\delta \phi _{\alpha }\right| n_{\mathbf{q}}\right\rangle \left\langle n_{%
\mathbf{k}}-1\left| \delta \phi _{\beta }\right|
n_{\mathbf{k}}\right\rangle.
\end{equation}
With the help of Eq.\ (\ref{angle2}) one obtains
\begin{equation}
M_{\mathrm{ph-R}}^{\alpha \beta } =\frac{\hbar ^{2}}{8\rho V}\frac{\left[ \mathbf{%
k}\times \mathbf{e}_{\mathbf{k}\lambda _{\mathbf{k}}}\right]
_{\alpha }\left[ \mathbf{q}\times \mathbf{e}_{\mathbf{q}\lambda
_{\mathbf{q}}}\right] _{\beta
}}{\sqrt{\hbar \omega _{\mathbf{k}\lambda _{\mathbf{k}}}\hbar \omega _{%
\mathbf{q}\lambda _{\mathbf{q}}}}}\sqrt{\left( n_{\mathbf{q}}+1\right) n_{%
\mathbf{k}}}.
\end{equation}
On the other hand, using the definition (\ref{MphRDef2}), one
obtains from Eq.\ (\ref{M11})
\begin{eqnarray}
&& M_{R}^{(1+1)} = \nonumber \\ &&-\sum_{\xi }\frac{\left\langle
\psi _{-}\left| \left[ {\mathcal{\hat{H}}}_{A},S_{\alpha }\right]
\right| \psi _{\xi }\right\rangle \left\langle \psi _{\xi }\left|
\left[ {\mathcal{\hat{H}}}_{A},S_{\beta }\right] \right| \psi
_{+}\right\rangle }{E_{+}+\hbar \omega _{\mathbf{k}}-E_{\xi }}M_{\mathrm{ph-R%
}}^{\alpha \beta}  \notag \\
&&- \sum_{\xi }\frac{\left\langle \psi _{-}\left| \left[
{\mathcal{\hat{H}}}_{A},S_{\alpha }\right] \right| \psi _{\xi
}\right\rangle \left\langle \psi _{\xi }\left| \left[
{\mathcal{\hat{H}}}_{A},S_{\beta }\right] \right| \psi
_{+}\right\rangle }{E_{+}-E_{\xi }-\hbar \omega _{\mathbf{q}}}M_{\mathrm{ph-R%
}}^{ \beta \alpha}\,, \nonumber \\  \label{MRaman11Calc}
\end{eqnarray}
where $\xi $ now labels intermediate states of the spin only. The
intermediate phonon states are
$\left|n_{\mathbf{k}}-1,n_{\mathbf{q} }\right\rangle $ in the
first term and $\left| n_{\mathbf{k}},n_{\mathbf{q}
}+1\right\rangle $ in the second term.\\

\subsection{Transition between eigenstates of $S_z$}

Consider for example the spin Hamiltonian
\begin{equation}\label{spin-Hamiltonian1}
{\mathcal{\hat{H}}}_S = {\mathcal{\hat{H}}}_A  +
{\mathcal{\hat{H}}}_{Z} = -DS_z^2 - g\mu_BH_zS_z
\end{equation}
that commutes with $S_z$. The exact energy states of this
Hamiltonian are the eigenstates of the $S_z$ operator, $S_z |m
\rangle = m |m \rangle $, with energies given by
\begin{equation}\label{E0}
E_m = -Dm^2 - g\mu_BH_zm\,.
\end{equation}
Let us study the general case of the spin transition $m_+
\rightarrow m_-$. From equations (\ref{MR2b}), (\ref{MphRDef}) and
(\ref{spin-Hamiltonian1}) one obtains
\begin{eqnarray}\label{MR2Sz}
&& M_R^{(2)} =  D \sum_{\alpha \beta}
\tilde{M}_{\mathrm{ph-R}}^{\alpha \beta} \times
\\  &&  \sum_{m_{\xi}}(m_+^2 + m_-^2 - 2m_{\xi}^2)  \langle
m_- | S_{\alpha} | m_{\xi}\rangle \langle m_{\xi}| S_{\beta} |m_+
\rangle \nonumber
  \,,
\end{eqnarray}
with $\tilde{M}_{\mathrm{ph-R}}^{\alpha \beta} = \frac{1}{2}\left(
M_{\mathrm{ph-R}}^{\alpha \beta}+M_{\mathrm{ph-R}}^{\beta \alpha}
\right)$. The summation on $m_{\xi}$ runs over all the spin
states. On the other hand, Eq.\ (\ref{MRaman11Calc}) results in
\begin{eqnarray}\label{MRaman11Sz}
M_R^{(1+1)} & = & \sum_{m_{\xi}}\frac{D^2(m_-^2 -
m_{\xi}^2)(m_+^2-m_{\xi}^2)}{D(m_{\xi}^2-m_+^2) +
g\mu_BH(m_{\xi}-m_+) + \hbar \omega_{\bf k}}\nonumber \\
&& \times \langle m_-|S_{\alpha} |m_{\xi}\rangle \langle m_{\xi}
|S_{\beta}| m_+ \rangle M_{\mathrm{ph-R}}^{\alpha \beta }
\nonumber \\ & + &\sum_{m_{\xi}}\frac{D^2(m_-^2 -
m_{\xi}^2)(m_+^2-m_{\xi}^2)}{D(m_{\xi}^2-m_+^2) +
g\mu_BH(m_{\xi}-m_+) - \hbar \omega_{\bf q}}\nonumber \\
&& \times  \langle m_-|S_{\alpha} |m_{\xi}\rangle \langle m_{\xi}
|S_{\beta}| m_+ \rangle M_{\mathrm{ph-R}}^{\beta \alpha }\,,
\end{eqnarray}
where summation over $\alpha$ and $\beta$ is assumed.

\subsubsection{Adjacent spin levels,  $m \rightarrow m\pm 1$}
\begin{figure}
\unitlength1cm
\begin{picture}(13,5.5)
\centerline{\psfig{file=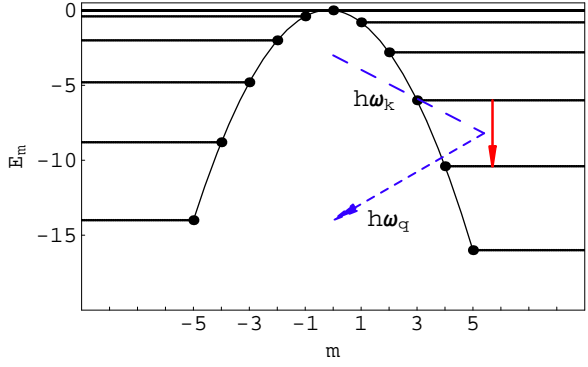,width=8cm}}
\end{picture}
\caption{\label{pRaman1} Transition between adjacent spin-energy
levels of Hamiltonian (\ref{spin-Hamiltonian1}) mediated by Raman
processes. }
\end{figure}

We will first treat the Raman processes involving transitions
between adjacent levels of the spin-Hamiltonian
(\ref{spin-Hamiltonian1}) (see Fig.\ \ref{pRaman1}). Therefore,
the spin-states in this case will be
\begin{equation}\label{spin-states1}
| m_+ \rangle = | m \rangle\,, \qquad | m_- \rangle = | m \pm 1
\rangle \,,
\end{equation}
where the plus sign applies to positive $m$ and the minus sign
applies to negative $m$. A straightforward calculation of the
matrix elements in equations (\ref{MR2Sz}) and (\ref{MRaman11Sz})
leads to
\begin{eqnarray}\label{MR2Sz1}
&& M_{R}^{(2)} = D(-2m \mp 1)l_{m, m\pm 1}
\tilde{M}_{\mathrm{ph-R}}^{\pm z}\nonumber \\
&& M_{R}^{(1+1)} =  0\,,
\end{eqnarray}
where
\begin{equation}\label{Mphzpm}
\tilde{M}_{\mathrm{ph-R}}^{z \pm }  =
\frac{1}{2}(\tilde{M}_{\mathrm{ph-R}}^{z x } \mp i
\tilde{M}_{\mathrm{ph-R}}^{z y })
\end{equation}
and $l_{m, m \pm 1} = \sqrt{(S \mp m)(S \pm m +1)}$.\\
Then $M_R = M_R^{(2)}$ and the transition rate is given by
\begin{equation}\label{Rate1}
\Gamma_{R}^{m\rightarrow m \pm 1} =
\sum_{{\scriptsize{\begin{matrix} {\bf k} \lambda_{\bf k}\\ {\bf
q} \lambda_{\bf q}
\end{matrix}}}} \frac{2\pi}{\hbar}
|M_{R}^{(2)}|^2 \delta\left( \hbar\omega_{\bf q} -
\hbar\omega_{\bf k} +E_{m\pm 1} - E_m \right)\,.
\end{equation}
Note that in the sums over the polarizations $\lambda_{\bf k}$ and
$\lambda_{\bf q}$, only the two transverse modes are considered.
To complete the calculation, we make use of
\begin{eqnarray}
[{\bf k}\times {\bf e}_{{\bf k} t}] & = & \pm k {\bf e}_{{\bf k}
t'}
\label{relation1}\\
\sum_{t = t_1, t_2}({\bf e}_{{\bf k}t}\cdot {\bf a})({\bf e}_{{\bf
k}t}\cdot {\bf b}) & = & {\bf a}\cdot {\bf b} - \frac{({\bf
k}\cdot {\bf a})({\bf k}\cdot{\bf b})}{k^2} \label{relation2}\,
\end{eqnarray}
and the replacement of $ \sum_{\bf k}$  by   $V\int d^3k/(2\pi)^3$
to obtain
\begin{equation}\label{ratefinalSz1}
\Gamma_R^{m\rightarrow m \pm 1}  =  \frac{1}{\hbar}\frac{l_{m,
m\pm1}^2I_{R1}}{\pi^3} \frac{ [D(-2m \mp 1)]^2}{{\cal{E}}^8_t}(k_B
T)^7\,,
\end{equation}
where
\begin{equation}\label{E-t}
{\cal{E}}_t \equiv (\hbar^3 \rho v_t^5)^{1/4}
\end{equation}
is a characteristic energy in the problem. In these expressions
$\rho$ is the mass density and $v_t$ is velocity of transverse
sound. $I_{R1}$ is given by
\begin{equation}\label{IR0}
I_{R1} = \frac{1}{1152}\int_0^{\theta_D/T} dx
x^3(x+\epsilon_1)^3\frac{e^{x+\epsilon_1}}{(e^x -
1)(e^{x+\epsilon_1}-1)}\,,
\end{equation}
where
\begin{equation}\label{epsilon1}
\epsilon_1 = \frac{E_m-E_{m\pm1}}{k_BT} = \frac{D(\pm 2m +1) \pm
g\mu_B H}{k_B T}\,.
\end{equation}
We remind the reader that in the above formulas the choice of
upper and lower sign corresponds the choice of $\pm$ in Eq.\
(\ref{spin-states1}). We use $m+1$ for positive $m$ and we use
$m-1$ for negative $m$, so that $E_m-E_{m\pm1}
> 0$.

\subsubsection{Non-adjacent spin levels,  $m \rightarrow m\pm 2$}
It is clear from equations (\ref{MR2Sz}) and (\ref{MRaman11Sz})
that the only allowed transitions between non-adjacent spin levels
are those with
\begin{equation}\label{transitionsSz2}
| m_+ \rangle = | m \rangle\,, \qquad | m_- \rangle = | m \pm 2
\rangle \,.
\end{equation}
\begin{figure}
\unitlength1cm
\begin{picture}(13,5.5)
\centerline{\psfig{file=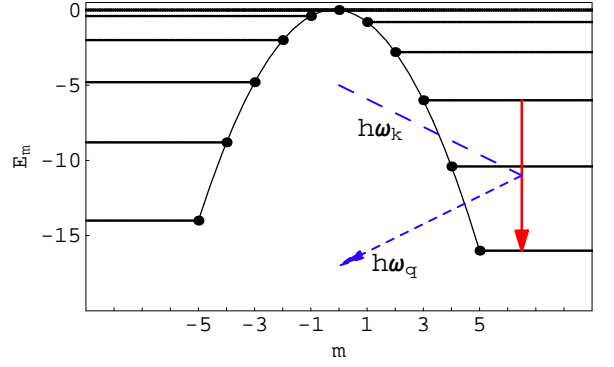,width=8cm}}
\end{picture}
\caption{\label{pRaman2} Transition between the spin-energy levels
of Hamiltonian (\ref{spin-Hamiltonian1}) $m \rightarrow m+2$ due
to Raman scattering. }
\end{figure}
In this case, represented in Fig.\ \ref{pRaman2}, equations
(\ref{MR2Sz}) and (\ref{MRaman11Sz}) give
\begin{eqnarray}\label{MR2Sz2}
&& M_R^{(2)}  =  D\,l_{m, m \pm 1}l_{m \pm 1, m \pm
2}M_{\mathrm{ph-R}}^{\pm \pm }\nonumber \\ && M_R^{(1+1)} =
-\frac{D^2}{2}(\pm 2m +3)(\pm2m +1)\,l_{m, m \pm 1}l_{m \pm 1, m
\pm 2} \nonumber \\ && \times \left(\frac{M_{\mathrm{ph-R}}^{\pm
\pm}}{E_{m\pm1} -E_m +\hbar \omega_{\bf k}}+
\frac{M_{\mathrm{ph-R}}^{\pm \pm}}{E_{m\pm1} -E_m - \hbar
\omega_{\bf q}}\right)\,, \nonumber \\
\end{eqnarray}
where
\begin{eqnarray}
M_{\mathrm{ph-R}}^{--}& = & \frac{1}{2}\left[M_{\mathrm{ph-R}}^{x
x } + i(M_{\mathrm{ph-R}}^{x y }+M_{\mathrm{ph-R}}^{y x}) -
M_{\mathrm{ph-R}}^{y y} \right]\nonumber \\
M_{\mathrm{ph-R}}^{++} & = & \frac{1}{2}\left[M_{\mathrm{ph-R}}^{x
x } - i(M_{\mathrm{ph-R}}^{x y }+M_{\mathrm{ph-R}}^{y x}) -
M_{\mathrm{ph-R}}^{y y} \right]\,.\nonumber \\
\end{eqnarray}
The transition rate can be obtained by computing
\begin{equation}\label{RateFormulaSz2}
\Gamma_{R}^{m\rightarrow m \pm 2} =
\sum_{{\scriptsize{\begin{matrix} {\bf k} \lambda_{\bf k}\\ {\bf
q} \lambda_{\bf q}
\end{matrix}}}} \frac{2\pi}{\hbar}|M_{R}|^2 \delta\left(
\hbar\omega_{\bf q} - \hbar\omega_{\bf k} +E_{m\pm 2} - E_m
\right)\,,
\end{equation}
with $M_R$ given by equations (\ref{M}) and (\ref{MR2Sz2}). Again,
we use equations (\ref{relation1}) and (\ref{relation2}) and the
replacement of $ \sum_{\bf k}$  by   $V d^3k/(2\pi)^3$ (with $V$
being the volume of the crystal) to obtain the final result
\begin{equation}\label{ratefinalSz2}
\Gamma_R^{m\rightarrow m \pm 2} = \frac{1}{\hbar}\frac{l_{m, m \pm
1}^2l_{m \pm 1, m \pm 2}^2\,I_{R2}}{\pi^3} \frac{
D^2}{{{\cal{E}}^8_t}}(k_B T)^7\,,
\end{equation}
where
\begin{eqnarray}\label{I1}
&& I_{R2} = \frac{1}{288}\int_0^{\theta_D/T}dx\frac{(x +
\epsilon_2)^3 x^3e^{(x+\epsilon_2)}}{[e^x -
1][e^{(x+\epsilon_2)}-1]} \bigg\{ 1 -  \nonumber
\\  && \frac{1}{2}(\pm 2m +1)(\pm 2m
+3)\left[\frac{D/k_BT}{\epsilon_1 +x} + \frac{D/k_BT}{\epsilon_1 -
\epsilon_2 - x} \right]
\bigg\}^2\nonumber \\
\end{eqnarray}
with
\begin{equation}\label{epsilon2}
\epsilon_2 = \frac{E_m - E_{m\pm2}}{k_BT} = \frac{4(\pm m +1) \pm2
g\mu_B H }{k_BT}\,.
\end{equation}
\subsection{Transitions between tunnel-split states}

Consider now a biaxial spin Hamiltonian with strong uniaxial
anisotropy
\begin{equation}\label{spin-Hamiltonian2}
{\mathcal{\hat{H}}}_S = {\mathcal{\hat{H}}}_A  - g\mu_B {\bf H}
\cdot {\bf S}\,, \qquad {\mathcal{\hat{H}}}_A = -DS_z^2 + E[S_x^2 -
S_y^2] \,,
\end{equation}
where $D \gg E>0$ and $DS \gg g\mu_B H_{\perp}$, with $H_{\perp}$
being the transverse magnetic field. Consequently,
${\mathcal{\hat{H}}}_A$ nearly commutes with $S_z$. The energy
levels of this Hamiltonian are approximately given by Eq.\
(\ref{E0}). The two levels $m$ and $m'$ are in resonance for the
values of the magnetic field $H_{z, mm'}^{(res)} =
(m+m')D/(g\mu_B)$. The level bias is given by
\begin{equation}\label{bias}
W \equiv E_m - E_m' =  g\mu_B\left(H_z - H_{z,
mm'}^{(res)}\right)(m'-m)\,.
\end{equation}
\begin{figure}
\unitlength1cm
\begin{picture}(11,5.5)
\centerline{\psfig{file=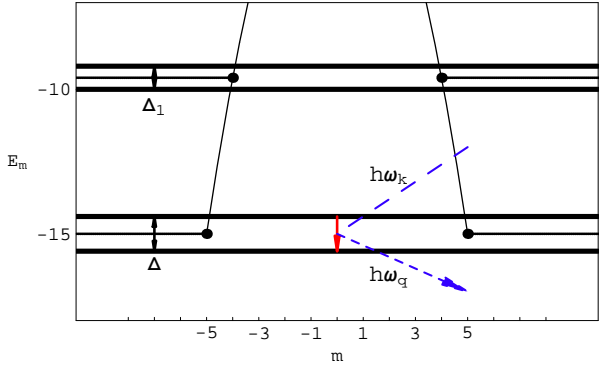,width=8cm}}
\end{picture}
\caption{\label{pRaman3} Transition between tunnel-split states
due to Raman scattering. }
\end{figure}
\subsubsection{The two state model}
Due to the terms in ${\mathcal{\hat{H}}}_S $ that do not commute
with $S_z$, the true eigenstates of ${\mathcal{\hat{H}}}_S $ far
from a resonance are given by expansions over the complete
$|m\rangle$ basis:
\begin{equation}\label{trueeigenstates}
|\phi_{m}\rangle = \sum _{m''=-S}^{S}c_{m m''}|m''\rangle\,,
\end{equation}
where $|c_{mm}|\simeq 1$ and all the other coefficients are small.
Hybridization of the states $|\phi_{m}\rangle$ and
$|\phi_{m'}\rangle$ when they are close to resonance can be taken
into account in the framework of the two-state model
\begin{eqnarray}\label{two-statemodel}
&& \left\langle \phi_{m_i} \left|{\mathcal{\hat{H}}}_S\right|
\phi_{m_i}\right\rangle = E_{m_i}, \quad m_i = m, m'\nonumber \\
&& \left\langle \phi_{m}\left|{\mathcal{\hat{H}}}_S\right|
\phi_{m'}\right\rangle = \frac{1}{2}\Delta\,,
\end{eqnarray}
where $\Delta$ is the tunnel splitting of the levels $m$ and $m'$
that can be calculated from the exact spin Hamiltonian
${\mathcal{\hat{H}}}_S $ \cite{Garanin} or determined
experimentally. Diagonalizing this $2\times2$ matrix yields the
eigenvalues
\begin{equation}\label{eigenvalues-twostate}
E_{\pm} = \frac{1}{2}\left(E_m + E_{m'} \pm \sqrt{W^2 + \Delta^2}
\right)\,.
\end{equation}
The corresponding eigenvectors can be represented in the form
\begin{equation}\label{eigenvectors-twostate}
|\psi_{\pm}\rangle = \frac{1}{\sqrt{2}}\left(C_{\pm}|\phi_{m}
\rangle \pm C_{\mp}|\phi_{m'}\rangle \right)\,,
\end{equation}
where
\begin{equation}\label{eigenvectors-twostatecoef}
C_{\pm} = \sqrt{1\pm \frac{W}{\sqrt{W^2 + \Delta^2}}}\,.
\end{equation}
Far from the resonance, $|W| \gg \Delta$, the eigenstates and
energy eigenvalues reduce to those of $|\phi_{m}\rangle$ and
$|\phi_{m'}\rangle$
states.\\

\subsubsection{Matrix elements}

Here we consider Raman processes involving spin transitions
between tunnel-split states (see Fig.\ \ref{pRaman3}). That is,
the spin eigenstates in Eq.\ (\ref{eigenstates}) are given by Eq.\
(\ref{eigenvectors-twostate}). In order to compute the matrix
element of the Raman process, we can rewrite $M_R^{(2)}$ by adding
and subtracting ${\mathcal{\hat{H}}}_Z$ in the spin matrix element
of Eq.\ (\ref{MR2b})
\begin{eqnarray}
&& \left\langle \psi _{-}\left| \left[ \left[
{\mathcal{\hat{H}}}_A,S_{\alpha }\right] ,S_{\beta }\right]
\right| \psi _{+}\right\rangle  = \left\langle \psi _{-}\left|
\left[ \left[ {\mathcal{\hat{H}}}_S,S_{\alpha }\right] ,S_{\beta
}\right]
\right| \psi _{+}\right\rangle  \notag \\
&& \qquad \qquad \qquad \qquad  -\left\langle \psi _{-}\left|
\left[ \left[ {\mathcal{\hat{H}}}_Z,S_{\alpha }\right] ,S_{\beta
}\right] \right| \psi _{+}\right\rangle \,.
\label{MatrElLabDefSubtr}
\end{eqnarray}
Taking into account that $|\psi_{\pm}\rangle$ are the eigenstates
of ${\mathcal{\hat{H}}}_S$ and inserting the identity $1 =
\sum_{\xi}|\psi_{\xi} \rangle \langle \psi_{\xi}|$ we can express
the first term of the right hand side as
\begin{eqnarray}\label{MatrElHSInserted}
&& \left\langle \psi _{-}\left| \left[ \left[
\hat{H}_{S},S_{\alpha }\right] ,S_{\beta }\right] \right| \psi
_{+}\right\rangle = \nonumber \\ && \sum_{\xi }\left(
E_{-}+E_{+}-2E_{\xi }\right) \left\langle \psi _{-}\left|
S_{\alpha }\right| \psi _{\xi }\right\rangle \left\langle \psi
_{\xi }\left| S_{\beta }\right| \psi _{+}\right\rangle. \qquad
\end{eqnarray}
Then,
\begin{eqnarray}
&& M_{R}^{(2)} = -\bigg\{ \sum_{\xi }\left( E_{-}+E_{+}-2E_{\xi
}\right) \left\langle \psi _{-}\left| S_{\alpha }\right| \psi
_{\xi }\right\rangle \times \notag \\
&& \left\langle \psi _{\xi }\left| S_{\beta }\right| \psi
_{+}\right\rangle \,
 - \left\langle \psi _{-}\left| \left[ \left[
{\mathcal{\hat{H}}}_Z,S_{\alpha }\right] ,S_{\beta }\right]
\right| \psi _{+}\right\rangle \bigg\}
\tilde{M}_{\mathrm{ph-R}}^{\alpha \beta}\,. \nonumber \\
\label{MRaman2Final}
\end{eqnarray}

Following the same procedure, we can rewrite $M_R^{(1+1)}$ from
Eq.\ (\ref{MRaman11Calc}) as an expansion on powers of $H$:
\begin{equation}\label{MR11twostate}
M_R^{(1+1)} = M_R^{(1+1)}(H^0) + M_R^{(1+1)}(H^1)+ O(H^2)
\end{equation}
with
\begin{eqnarray}
&& M_{R}^{(1+1)}(H^0) = \nonumber \\ && -\sum_{\xi }\left(
E_{-}-E_{\xi }\right) \left( E_{\xi }-E_{+}\right) \left\langle
\psi _{-}\left| S_{\alpha }\right| \psi _{\xi }\right\rangle
\left\langle \psi _{\xi }\left| S_{\beta }\right| \psi
_{+}\right\rangle
\notag \\
&&\times \left[ \frac{M_{\mathrm{ph-R}}^{\alpha \beta }}{%
E_{+}+\hbar \omega _{\mathbf{k}}-E_{\xi }}+\frac{M_{\mathrm{ph-R}%
}^{ \beta \alpha }}{E_{+}-E_{\xi }-\hbar \omega
_{\mathbf{q}}}\right] \label{MRaman11Calc2}
\end{eqnarray}
and
\begin{eqnarray}
&& M_{R}^{(1+1)}(H^1) = \nonumber \\ &&\sum_{\xi }\frac{\left\langle \psi _{-}\left| \left[ \hat{H}%
_{Z},S_{\alpha }\right] \right| \psi _{\xi }\right\rangle
\left\langle \psi _{\xi }\left| \left[ \hat{H}_{S},S_{\beta
}\right] \right| \psi
_{+}\right\rangle }{E_{+}+\hbar \omega _{\mathbf{k}}-E_{\xi }}M_{\mathrm{ph-R%
}}^{\alpha \beta }  \notag \\
&&+\sum_{\xi }\frac{\left\langle \psi _{-}\left| \left[ \hat{H}%
_{S},S_{\alpha }\right] \right| \psi _{\xi }\right\rangle
\left\langle \psi _{\xi }\left| \left[ \hat{H}_{Z},S_{\beta
}\right] \right| \psi
_{+}\right\rangle }{E_{+}+\hbar \omega _{\mathbf{k}}-E_{\xi }}M_{\mathrm{ph-R%
}}^{\alpha \beta }  \notag \\
&&+\sum_{\xi }\frac{\left\langle \psi _{-}\left| \left[ \hat{H}%
_{Z},S_{\alpha }\right] \right| \psi _{\xi }\right\rangle
\left\langle \psi _{\xi }\left| \left[ \hat{H}_{S},S_{\beta
}\right] \right| \psi
_{+}\right\rangle }{E_{+}-E_{\xi }-\hbar \omega _{\mathbf{q}}}M_{\mathrm{ph-R%
}}^{\beta \alpha }  \notag \\
&&+\sum_{\xi }\frac{\left\langle \psi _{-}\left| \left[ \hat{H}%
_{S},S_{\alpha }\right] \right| \psi _{\xi }\right\rangle
\left\langle \psi _{\xi }\left| \left[ \hat{H}_{Z},S_{\beta
}\right] \right| \psi
_{+}\right\rangle }{E_{+}-E_{\xi }-\hbar \omega _{\mathbf{q}}}M_{\mathrm{ph-R%
}}^{\beta \alpha }. \nonumber \\ \label{MRam11MixedDef}
\end{eqnarray}

\subsubsection{Transition rate for $H=0$}
At $ H = 0$ one obtains
\begin{eqnarray}
&& M_{R} =  -\sum_{\xi }\left\langle \psi _{-}\left| S_{\alpha
}\right| \psi _{\xi }\right\rangle \left\langle \psi _{\xi }\left|
S_{\beta }\right| \psi
_{+}\right\rangle  \left[ M_{\mathrm{ph-R}}^{(\alpha \beta )}\times \right. \notag \\
&&\left. \left( \frac{%
E_{-}+E_{+}}{2}-E_{\xi }+\frac{\left( E_{-}-E_{\xi }\right) \left(
E_{\xi }-E_{+}\right) }{E_{+}+\hbar \omega _{\mathbf{k}}-E_{\xi
}}\right) + \right.
\notag \\
&&\left.M_{\mathrm{ph-R}}^{( \beta \alpha)} \left( \frac{E_{-}+E_{+}}{2}%
-E_{\xi }+\frac{\left( E_{-}-E_{\xi }\right) \left( E_{\xi }-E_{+}\right) }{%
E_{+}-E_{\xi }-\hbar \omega _{\mathbf{q}}}\right) \right]
\nonumber \\ \label{MRFinalGen}
\end{eqnarray}
It is convenient to consider the terms with $\xi = \pm$ and $\xi
\neq \pm$ separately. The contribution from $\xi = \pm$ is
\begin{eqnarray}
&& M_{R}= -\left( E_{+}-E_{-}\right) \bigg( \left\langle \psi
_{-}\left| S_{\alpha }\right| \psi _{-}\right\rangle \left\langle
\psi _{-}\left| S_{\beta }\right| \psi _{+}\right\rangle \nonumber
\\ && -\left\langle \psi _{-}\left| S_{\alpha }\right| \psi
_{+}\right\rangle \left\langle \psi _{+}\left| S_{\beta }\right|
\psi _{+}\right\rangle \bigg) \tilde{M}_{\mathrm{ph-R}}^{\alpha
\beta }. \label{MRamanxipm}
\end{eqnarray}
Using the time-reversal symmetry, we obtain
\begin{equation}\label{time-reversal}
\langle \psi_{\pm} | {\bf S} |\psi_{\pm} \rangle = -\langle
\psi_{\pm} | {\bf S} |\psi_{\pm} \rangle ^{*}\,.
\end{equation}
For the biaxial model with $E > 0$ the states $|\psi_{\pm}\rangle$
are real. Then,
\begin{equation}
\langle \psi_{\pm} | S_z |\psi_{\pm} \rangle = \langle \psi_{\pm}
|  S_x |\psi_{\pm} \rangle = 0\,.
\end{equation}
On the other hand, $\langle \psi_{\pm} | S_y |\psi_{\pm}
\rangle=0$ because of the factorization of the Hilbert space:
$\mathcal{H}=\mathcal{H} _{1}^{(1)}\otimes \mathcal{H}_{1}^{(2)},$
$\left( -S,-S+2,\ldots ,S\right) \in \mathcal{H }_{1}^{(1)}$ and
$\left( -S+1,-S+3,\ldots ,S-1\right) \in \mathcal{H}_{1}^{(2)}$.
Thus Eq.\ (\ref{MRamanxipm}) yields a
zero result.\\

Let us consider now the terms with $\xi \neq \pm$. In this case,
the difference between the energies of the doublet is much smaller
than the energy distance to the other states,  $|E_{+} - E_-| \ll
|E_{\pm} - E_{\xi}|$, so that one can replace $E_{\pm }$ with $E$
and $\hbar \omega_{\mathbf{q}}$ with $\hbar \omega _{\mathbf{k}}$
in the matrix elements. We consider the case of low temperature,
when $ \hbar \omega _{\mathbf{k}}\ll \left| E-E_{\xi }\right| $
for thermal phonons. Then $M_{R}$ can be simplified to
\begin{equation}\label{MRtwostate}
M_{R} \simeq 2\left( \hbar \omega _{\mathbf{k}}\right)
^{2}\tilde{M}_{\mathrm{ph-R}}^{\alpha \beta } A_{\alpha \beta}\,,
\end{equation}
with
\begin{equation}
A_{\alpha \beta }\equiv \sum_{\xi }{}^{^{\prime
}}\frac{\left\langle \psi _{-}\left| S_{\alpha }\right| \psi _{\xi
}\right\rangle \left\langle \psi _{\xi }\left| S_{\beta }\right|
\psi _{+}\right\rangle }{E-E_{\xi }},\label{AalphabetaDef}
\end{equation}
where prime means that $\xi =\pm $ have been excluded. Note that
the quadratic dependence on $\omega_{\bf q}$ in Eq.\
(\ref{MRtwostate}) results from cancellations between terms from
$M_R^{(2)}$ and terms from $M_R^{(1+1)}$. Consequently, the
relaxation rate will have a different temperature dependence from
the result that one would obtain if one added the rates stemming
from $M_R^{(2)}$ and $M_R^{(1+1)}$ independently. The rate is
given by
\begin{eqnarray}
&& \Gamma_{R0}^{+\rightarrow -} = \sum_{{\bf k}, \lambda_{\bf
k}}\sum_{{\bf q}, \lambda_{\bf q}} \frac{2\pi}{\hbar} |M_{R}|^2
\delta\left( \hbar\omega_{\bf q} - \hbar\omega_{\bf k} \right)
\nonumber
\\ && = \frac{1}{18\hbar \left( 2\pi \right) ^{3}}\sum_{\alpha \beta }A_{\alpha \beta }\left(
A_{\alpha \beta}^{*} + A_{\beta \alpha}^{*}
\right)\frac{(k_BT)^{11}}{{{\cal{E}}^8_t}}I_{10}\nonumber \\
\label{WRtwostates2}
\end{eqnarray}
where equations (\ref{relation1}) and (\ref{relation2}) have been
used and the continuum limit has been taken. The constant $I_{10}$
is
\begin{equation}\label{I10}
I_{10} = \int_{0}^{\infty }dx\,\frac{x^{10}e^{x}}{\left(
e^{x}-1\right) ^{2}} = \Gamma (11)\zeta (10).
\end{equation}
For transitions between the lowest doublet $m=-S, m'=S$ specified
in Eq.\ (\ref{eigenvectors-twostate}) one can evaluate $ A_{\alpha
\beta }$ by considering also the first excited doublet $m=-S+1,
m'=S-1$
\begin{equation}\label{eigenvectors-twostate2}
\left| \psi _{1\pm }\right\rangle =\frac{1}{\sqrt{2}}\left(
C_{1\pm }\left|\phi_{m +1}\right\rangle \pm C_{1\mp }\left|
\phi_{{m}^{\prime }-1}\right\rangle \right)
\end{equation}
with $C_{1\pm }$ given by Eq.\ (\ref{eigenvectors-twostatecoef})
with $\Delta \rightarrow \Delta _{1}$ and $W\rightarrow W_{1}$.
Hence the only non zero matrix elements are
\begin{equation}
A_{\pm \mp}=\frac{1}{E_{m}-E_{m+1}}\bar{A}_{\pm \mp},  \label{Apm}
\end{equation}
where
\begin{equation}
\bar{A}_{\pm \mp}\equiv \sum_{\eta = \pm }\left\langle \psi
_{-}\left| S_{\pm}\right| \psi _{1\eta }\right\rangle \left\langle
\psi _{1\eta }\left| S_{\mp}\right| \psi _{+}\right\rangle .
\label{Abarpm}
\end{equation}
Evaluation of these matrix elements yields
\begin{equation}\label{Aalphabeta}
\sum_{\alpha \beta }A_{\alpha \beta }\left( A_{\alpha \beta}^{*} +
A_{\beta \alpha}^{*} \right) = \frac{S^2}{\left(E_{-S} -
E_{-S+1}\right)^2} \frac{\Delta ^{2}}{W^{2}+\Delta ^{2}}\,.
\end{equation}
For the case under consideration $W =0$ and the transition rate is,
then
\begin{equation}
\Gamma_{R0}^{+\rightarrow -}=\frac{80\,\pi^7S^2}{297\hbar} \frac{
(k_BT)^{11}}{\left( E_{-S}-E_{-S+1}\right) ^{2}{{\cal{E}}^8_t}}.
\label{WRnoHFinal}
\end{equation}

\subsubsection{Transition rate in the presence of a magnetic field}
Here we are going to evaluate the contribution of the magnetic
field ${\bf H}$ to the rate of the transition between the
tunnel-split ground-state levels. We consider magnetic fields with
very small longitudinal component, $g\mu_B H_z \sim \Delta$. For
longitudinal fields $g \mu_B H_z \gg \Delta$ the Raman processes
die out. We are also restricted to not very large transverse
magnetic fields, $g\mu_B H_{\perp} \ll DS$. In this case, it is
sufficient to compute the lowest-order contribution of the
magnetic field to the transition rate. This contribution can be
essential because of the cancellation that occurs in the matrix
element for the $H = 0$ case, Eq.\ (\ref{MRtwostate}). This
cancellation leads to the $\Gamma_{R}^{+\rightarrow -} \propto
T^{11}$ dependence. As we shall see, there is no such cancellation
in the field-dependent term, so that the result shows a $
T^{7}$-dependence.

According to Eq.\ (\ref{MRaman2Final}), the linear order
contribution of $H$ to $M_R^{(2)}$ is given by
\begin{equation}\label{MR2linearH}
\delta M_{R}^{(2)} =\left\langle \psi _{-}\left| \left[ \left[ \hat{H}%
_{Z},S_{\alpha }\right] ,S_{\beta }\right] \right| \psi
_{+}\right\rangle  \tilde{M}_{\mathrm{ph-R}}^{\alpha \beta }\,.
\end{equation}
The double commutator equals
\begin{equation}\label{Hzcommutator}
\left[ \left[ \hat{H}_{Z},S_{\alpha }\right] ,S_{\beta }\right] =
g \mu _{B}\left( H_{\beta }S_{\alpha }-H_{\gamma }S_{\gamma
}\delta _{\alpha \beta }\right)
\end{equation}
so that
\begin{eqnarray}\label{MR2linearH2}
\delta M_{R}^{(2)} & = & g\mu _{B}\left( H_{\beta }\delta_{\alpha
z}-H_{z}\delta _{\alpha \beta
}\right)\tilde{M}_{\mathrm{ph-R}}^{\alpha \beta } \left\langle
\psi _{-}\left| S_{z}\right| \psi _{+}\right\rangle \nonumber \\
& = & g\mu _{B}K_{\alpha \beta
}^{(2)}\tilde{M}_{\mathrm{ph-R}}^{\alpha \beta },
\end{eqnarray}
where
\begin{equation}
K_{\alpha \beta }^{(2)}=-\left( H_{\beta }\delta _{\alpha
z}-H_{z}\delta
_{\alpha \beta }\right) \frac{\Delta }{\sqrt{W^{2}+\Delta ^{2}}}\frac{%
m^{\prime }-m}{2}.  \label{Kalbet2Def}
\end{equation}
The phonon matrix element $\tilde{M}_{\mathrm{ph-R}}^{\alpha \beta
}$ is symmetric in $\alpha \beta$. Thus, it is possible and
convenient to replace $K_{\alpha \beta }^{(2)}$ in Eq.\
(\ref{MR2linearH2}) by the symmetrized tensor
\begin{equation}\label{symmetrization}
\tilde{K}_{\alpha \beta}^{(2)} = \frac{1}{2}\left(K_{\alpha \beta
}^{(2)} + K_{ \beta \alpha }^{(2)}  \right)\,.
\end{equation}
On the other hand, the linear order contribution of $H$ to
$M_R^{(1+1)}$ is given by Eq.\ (\ref{MRam11MixedDef}). By using
Eq.\ (\ref{Hzcommutator}) and assuming $\hbar \omega_{\bf k} \ll
E_{m+1}-E_m$ we obtain
\begin{eqnarray}
\delta M_{R}^{(1+1)} &\cong & ig\mu _{B}2H_{\gamma }\left( \epsilon
_{\gamma \alpha \delta }\bar{A} _{\delta \beta }-\epsilon _{\gamma
\beta \delta }\bar{A}_{\alpha \delta }\right)
\tilde{M}_{\mathrm{ph-R}}^{\alpha \beta }\nonumber
\\ & \equiv & g\mu _{B}K_{\alpha \beta
}^{(1+1)}\tilde{M}_{\mathrm{ph-R}}^{\alpha \beta },
\label{MRam1p1K}
\end{eqnarray}
where
\begin{equation}
K_{\alpha \beta }^{(1+1)}=i2H_{\gamma }\left( \epsilon _{\gamma
\alpha \delta }\bar{A}_{\delta \beta }-\epsilon _{\gamma \beta
\delta }\bar{A}_{\alpha \delta }\right) . \label{Kalbet1p1Def}
\end{equation}
Again, it is convenient to replace $K_{\alpha \beta }^{(1+1)}$ in
Eq.\ (\ref{MRam1p1K}) by the symmetrized version
\begin{equation}\label{symmetrization2}
\tilde{K}_{\alpha \beta}^{(1+1)} = \frac{1}{2}\left(K_{\alpha
\beta }^{(1+1)} + K_{ \beta \alpha }^{(1+1)}  \right)\,.
\end{equation}
The addition to the Raman matrix element due to $H \neq 0$ is,
then
\begin{equation}\label{MRtwo-stateH}
\delta M_R \equiv \delta M_R ^{(2)} + \delta M_R ^{(1+1)} =
g\mu_B\tilde{K}_{\alpha \beta}\tilde{M}_{\mathrm{ph-R}}^{\alpha
\beta }\,,
\end{equation}
with $\tilde{K}_{\alpha \beta} = \tilde{K}_{\alpha
\beta}^{(2)}+\tilde{K}_{\alpha \beta}^{(1+1)}$. The transition
rate is based upon $|M_R + \delta M_R|^2$. Here, $|M_R|^2$ was
taken into account above. The interference term can be shown to be
proportional to $H_z$ and thus negligibly small. Therefore, the
field effect is entirely contained in the term $|\delta M_R|^2$.
Applying the same procedure as in the previous section, one
obtains the following addition to the relaxation rate
\begin{equation}
\delta \Gamma_{R}^{+ \rightarrow -}=\frac{1}{\hbar}\frac{\pi^3}{
3024}\left( g\mu _{B}\right) ^{2}\sum_{\alpha \beta }|
\tilde{K}_{\alpha \beta }| ^{2}\frac{(k_BT)^{7}}{{{\cal{E}}^8_t}}.
\label{WRField}
\end{equation}
Evaluation of $\sum_{\alpha \beta }| \tilde{K}_{\alpha \beta }|
^{2}$ for the lowest doublet $m = -S, m'=S$ yields
\begin{equation}\label{sumK2}
\sum_{\alpha \beta }| \tilde{K}_{\alpha \beta }| ^{2} = 2\left(
H_{z}^{2}+\frac{1}{4}H_\perp^2\right) \frac{\Delta ^{2}}{
W^{2}+\Delta ^{2}}\,S^2\,.
\end{equation}
Therefore, the transition rate in the presence of a magnetic field
is
\begin{equation}\label{Rate-twostate-H}
\Gamma_R^{+ \rightarrow -}({\bf H}) \simeq \frac{ \Delta
^{2}}{W^{2}+\Delta ^{2}}\, \left(\Gamma_{R0}^{+ \rightarrow -} +
\delta\Gamma_{R0}^{+ \rightarrow -}\right)\,,
\end{equation}
where $\Gamma_{R0}^{+ \rightarrow -}$ is given by Eq.\
(\ref{WRnoHFinal}) and
\begin{equation}
\delta \Gamma_{R0}^{+ \rightarrow -} = \frac{\pi^3S^2}{6048\hbar}
\frac{H_{\perp}^2\left( g\mu _{B}\right) ^{2}
(k_BT)^{7}}{{{\cal{E}}^8_t}}\,. \label{WRFieldfinal}
\end{equation}
Note that we have used $H_z \ll H_{\perp}$ in the last expression.
One can see from Eq.\ (\ref{Rate-twostate-H}) that the relaxation
rate due to Raman processes dies out when going out of resonance,
$W \gg \Delta$.

\section{Processes involving emission of two phonons}
For processes involving emission of two phonons of, say, wave
vectors $\mathbf{k}$ and $\mathbf{q}$ we use the following
designations:
\begin{equation}
\left| \phi _{+}\right\rangle \equiv \left| n_{\mathbf{k}},n_{\mathbf{q}%
}\right\rangle ,\qquad \left| \phi _{-}\right\rangle \equiv \left| n_{%
\mathbf{k}}+1,n_{\mathbf{q}}+1\right\rangle  \label{phiviankEm}
\end{equation}
In this case, conservation of energy reads:
\begin{equation}
E_{+}- E_-=\hbar \omega_{\bf k} + \hbar \omega_{\bf q}.
\label{EnergyconservationE2}
\end{equation}

The matrix element for this process is, again, the sum of the
matrix element
with $\hat{{\mathcal H}}_{\mathrm{s-ph}}^{(2)}$ and that with $\hat{{\mathcal H}}_{\mathrm{s-ph}%
}^{(1)}$ in the second order:
\begin{equation}
M_{E}=M_{E}^{(2)}+M_{E}^{(1+1)},  \label{ME}
\end{equation}
where according to equations (\ref{Hsph}) and (\ref{M2}),
\begin{eqnarray}
&& M_{E}^{(2)} =-\frac{1}{2}\left\langle \psi _{-}\left| \left[
\left[ \hat{{\mathcal H}}_{A},S_{\alpha }\right] ,S_{\beta
}\right] \right| \psi _{+}\right\rangle
\notag \\
&& \qquad \times \left\langle
n_{\mathbf{k}}-1,n_{\mathbf{q}}+1\left| \delta
\phi _{\alpha }\delta \phi _{\beta }\right| n_{\mathbf{k}},n_{\mathbf{q}%
}\right\rangle .  \label{MatrElLabDefE}
\end{eqnarray}
In this case it is convenient to express the phonon matrix element
as:
\begin{eqnarray}
&&\left\langle n_{\mathbf{k}}+1,n_{\mathbf{q}}+1\left| \delta \phi
_{\alpha }\delta \phi _{\beta }\right|
n_{\mathbf{k}},n_{\mathbf{q}}\right\rangle \nonumber \\ && = M_{
\mathrm{ph-E}}^{\alpha \beta }+M_{\mathrm{ph-E}}^{\beta \alpha
}\equiv 2\tilde{M}_{\mathrm{ph-E}}^{\alpha \beta},
\label{MphRDefE}
\end{eqnarray}
where
\begin{eqnarray}\label{MphE}
&& M_{\mathrm{ph-E}}^{\alpha \beta } = \left\langle
n_{\mathbf{q}}+1\left| \delta \phi _{\alpha }\right|
n_{\mathbf{q}}\right\rangle \left\langle n_{ \mathbf{k}}+1\left|
\delta \phi _{\beta }\right| n_{\mathbf{k}}\right\rangle \nonumber
\\ && = \frac{\hbar ^{2}}{8\rho V}\frac{\left[ \mathbf{ k}\times
\mathbf{e}_{\mathbf{k}\lambda _{\mathbf{k}}}\right] _{\alpha
}\left[ \mathbf{q}\times \mathbf{e}_{\mathbf{q}\lambda
_{\mathbf{q}}}\right] _{\beta }}{\sqrt{\hbar \omega
_{\mathbf{k}\lambda _{\mathbf{k}}}\hbar \omega _{
\mathbf{q}\lambda _{\mathbf{q}}}}}\sqrt{\left(
n_{\mathbf{q}}+1\right) (n_{
\mathbf{k}}+1)}.\nonumber \\
\end{eqnarray}

\subsection{Transitions between eigenstates of $S_z$}
As stated above, the spin-phonon relaxation by emission of two
phonons may be more important than the relaxation by Raman
processes only if the energy difference between the spin-states
satisfies $\Delta E \ll k_B T$. Provided that the energy
difference between tunnel-split levels, $\Delta$, is very small,
only the relaxation by the emission of two phonons between
eigenstates of $Sz$ will be considered.

To this end, we will make use of the spin-Hamiltonian
(\ref{spin-Hamiltonian1}) and the transitions between its
eigenstates, $|m\rangle$.
\subsubsection{Adjacent spin levels, $m \rightarrow m\pm 1$}
The matrix elements in this case are
\begin{eqnarray}\label{MEadj}
&& M_{E}^{(2)}  =    D(\mp
2m+1)l_{m,m\pm1}\tilde{M}_{\mathrm{ph-E}}^{z \pm }
\nonumber \\
&& M_{E}^{(1+1)} =  0\,,
\end{eqnarray}
with
\begin{equation}
\tilde{M}_{\mathrm{ph-E}}^{z \pm }=
\frac{1}{2}\left[\tilde{M}_{\mathrm{ph-E}}^{z x } \mp
i\tilde{M}_{\mathrm{ph-E}}^{z y }\right]\,.
\end{equation}
The decay rate is then given by
\begin{equation}\label{RateEadj}
\Gamma_{E}^{m \rightarrow m\pm 1} =
\sum_{{\scriptsize{\begin{matrix} {\bf k} \lambda_{\bf k}\\ {\bf
q} \lambda_{\bf q}
\end{matrix}}}}\frac{2\pi}{\hbar} |M_{E}^{(2)}|^2
\delta\left(\hbar \omega_{\bf k} + \hbar \omega_{\bf q} +E_{m\pm
1} - E_m) \right).
\end{equation}
Using the same techniques as in the previous calculations, one
obtains
\begin{equation}\label{RateEadjfinal}
\Gamma_{E}^{m \rightarrow m\pm 1} =
\frac{1}{\hbar}\frac{l_{m,m\pm1}^2
I_{E1}}{\pi^3}\frac{\left[D(2m\pm
1)\right]^2}{{{\cal{E}}^8_t}}(k_BT)^7\,,
\end{equation}
where
\begin{equation}\label{IE1}
I_{E1} = \frac{1}{1152}\int_0^{\epsilon_1}dx \frac{x^3 (\epsilon_1
-x)^3e^{\epsilon_1}}{(e^x -1)(e^{(\epsilon_1 -x)} -1)}\,.
\end{equation}

\subsubsection{Non-adjacent spin levels, $m \rightarrow m \pm 2$}
In this case, the matrix elements are
\begin{eqnarray}\label{MEnonadj}
&& M_{E}^{(2)} = D\, l_{m, m \pm 1} l_{m \pm 1, m \pm
2}{M}_{\mathrm{ph-E}}^{\pm \pm
}\nonumber \\
 && M_E^{(1+1)} =
-\frac{D^2}{2}(\pm 2m +3)(\pm2m +1)\,l_{m, m \pm 1}l_{m \pm 1, m
\pm 2} \nonumber \\ && \times \left(\frac{M_{\mathrm{ph-E}}^{\pm
\pm}}{E_{m\pm1} -E_m +\hbar \omega_{\bf k}}+
\frac{M_{\mathrm{ph-E}}^{\pm \pm}}{E_{m\pm1} -E_m - \hbar
\omega_{\bf q}}\right)\,, \nonumber \\
\end{eqnarray}
where
\begin{equation}
{M}_{\mathrm{ph-E}}^{\pm \pm } =
\frac{1}{2}\left[{M}_{\mathrm{ph-E}}^{x x} -
{M}_{\mathrm{ph-E}}^{y y} \mp i({M}_{\mathrm{ph-E}}^{x
y}+{M}_{\mathrm{ph-E}}^{y x} ) \right]\,.
\end{equation}
The decay rate is
\begin{equation}\label{RateEnonadj}
\Gamma_{E}^{m \rightarrow m\pm 2} =\frac{1}{\hbar}\frac{l_{m, m
\pm 1}^2 l_{m\pm1, m\pm 2}^2 I_{E2}}{\pi^3}\frac{D^2 (k_B
T)^7}{{{\cal{E}}^8_t}}\,,
\end{equation}
with

\begin{eqnarray}
&& I_{E2} = \frac{1}{288}\int_0^{\epsilon_2}dx\frac{(\epsilon_2 -
x)^3 x^3e^{\epsilon_2}}{[e^x - 1][e^{(\epsilon_2-x)}-1]} \bigg\{ 1
- \nonumber
\\  && \frac{1}{2}(\pm 2m +1)(\pm 2m
+3)\left[\frac{D/k_BT}{\epsilon_1 +x} + \frac{D/k_BT}{\epsilon_1 +
\epsilon_2 - x} \right]
\bigg\}^2\,.\nonumber \\
\end{eqnarray}

\section{Discussion}

Let us now compare transition rates obtained for direct, or
one-phonon, processes \cite{CGS, Book} and the rates of the same
transitions due to Raman processes.

For spin transitions between adjacent eigenstates of the spin
Hamiltonian (\ref{spin-Hamiltonian1}), the ratio of the Raman rate
and the direct rate is given by
\begin{equation}\label{comparison1}
\frac{\Gamma_R^{m\rightarrow m+1}}{\Gamma_D^{m\rightarrow m+1}} =
\frac{24 I_{R1}}{\pi^2} \frac{(k_BT)^7 [1- e^{-{\Delta
E}/({k_BT})}]}{{\cal{E}}_t^4(\Delta E)^3}\,,
\end{equation}
\\
where $\Delta E \equiv E_m - E_{m+1}$. In the limit of $k_BT \ll
\Delta E < \theta_D$ transitions are dominated by direct
processes. When $\Delta E \ll k_BT < \theta_D$ one has
\begin{equation}
\frac{\Gamma_R^{m\rightarrow m+1}}{\Gamma_D^{m\rightarrow m+1}}
\sim \left(\frac{{\cal{E}}_t}{\Delta E}\right)^2
\left(\frac{k_BT}{{\cal{E}}_t}\right)^6\,.
\end{equation}
In this case transitions can be easily dominated by the Raman
processes. For example, for $S = 10$, $m =5$, $D = 0.1$K, $g\mu_B
H = 1K$ the energy difference is $\Delta E = 2.1$K, so that at $ T
= 35K$ the rate of the direct process is $\Gamma_D^{m\rightarrow
m+1} \simeq 10^5s^{-1}$, while the rate of the Raman process is $
\Gamma_R^{m\rightarrow m+1} \simeq 10^6 s^{-1}$. In these
estimates we used the value of ${\cal{E}}_t \sim 100$K. Note that
in such molecular magnets as Mn-12 and Fe-8 it will be difficult
to have the corresponding Raman processes dominant because of
large distances between adjacent spin levels and small Debye
temperature. Processes involving the emission of two phonons
cannot be dominant in any temperature range. In some range they
can dominate over Raman processes but not over direct processes.

For spin transitions between tunnel-split states of the spin
Hamiltonian (\ref{spin-Hamiltonian2}) at $ H =0$, the ratio of the
Raman rate over the direct rate at $\Delta \ll k_BT$ is given by
\begin{equation}\label{comparison2}
\frac{\Gamma_{R0}^{+\rightarrow -}}{\Gamma_D^{+\rightarrow -}} =
\frac{960
\pi^8}{297(2S-1)^2}\;\left(\frac{k_BT}{D}\right)^2\left(\frac{k_BT}{{\cal{E}}_t}\right)^4
\left(\frac{k_BT}{\Delta}\right)^{4}\,.
\end{equation}
Consequently, in zero field and temperatures significantly
exceeding $\Delta$, the Raman processes will have much higher
probability than direct processes. At, e.g., $D = 0.5$K, $S = 10$,
$ \Delta = 10^{-2}$K and $T=5$K, the rate of the direct process
gives $\Gamma_D^{+\rightarrow -} \simeq 10^{-4}s^{-1}$, while the
rate of the Raman process will be $\Gamma_R^{+\rightarrow -}
\simeq 5\cdot10^4 s^{-1}$. Note that at temperatures where Raman
processes dominate over direct processes, contribution of the
magnetic field to the rate, Eq.\ (\ref{WRFieldfinal}), is small
compared to the zero-field rate, Eq.\ (\ref{WRnoHFinal}).

CC thanks Jaroslav Albert for useful discussions. This work has
been supported by the NSF Grant No. EIA-0310517.

\end{document}